 \documentclass[final,5p,times,sort,compress]{elsarticle}

\usepackage{amssymb}
\usepackage{lipsum}
\usepackage{amsthm}
\usepackage{multirow}

\usepackage{xcolor}
\colorlet{mycyan}{cyan!70!black}
\usepackage[
    colorlinks=true,
    linkcolor=mycyan,
    citecolor=mycyan,
    urlcolor=mycyan,
    pdfborder={0 0 0},
    breaklinks=true
]{hyperref}

\makeatletter

\renewenvironment{thebibliography}[1]
  {\begin{oldbib}{#1}%
   \small
   \setlength{\itemsep}{0pt}%
   \setlength{\parskip}{0pt}%
   \setlength{\parsep}{0pt}%
   \setlength{\topsep}{2pt}}
  {\end{oldbib}}
\makeatother

\newcommand{\OOB}{\Omega^-\bar{\Omega}^+}

\newcommand{\EE}{e^+e^-}
\newcommand{\BB}{B\bar{B}}

\newcommand{\psp}{\psi(3686)}
\newcommand{\jpsi}{J/\psi}
\newcommand{\ar}{\rightarrow}

\journal{Physics Letters B}

\AtBeginDocument{
    \hypersetup{
        linkcolor=mycyan,
        citecolor=mycyan,
        urlcolor=mycyan
    }
}

\begin{document}

\begin{frontmatter}

\title{Search for Charmonium(-like) states decaying into the $\Omega^-\bar\Omega^+$ final states}

\author[a,b]{Ruoyu Zhang}
\author[a,b,c]{Xiongfei Wang}
\ead{wangxiongfei@lzu.edu.cn}
\affiliation[a]{organization={School of Physical Science and Technology},
            addressline={Lanzhou University}, 
            city={Lanzhou},
            postcode={730000}, 
            state={},
            country={China}}
\affiliation[b]{organization={Lanzhou Center for Theoretical Physics, Key Laboratory of Theoretical Physics of Gansu Province},
            addressline={Gansu Provincial Research Center for Basic Disciplines of Quantum Physics}, 
            city={Lanzhou},
            postcode={730000}, 
            state={},
            country={China}}
\affiliation[c]{organization={Frontiers Science Center for Rare Isotopes},
            addressline={Lanzhou University}, 
            city={Lanzhou},
            postcode={730000}, 
            state={},
            country={China}}

\begin{abstract}
Recently, the BESIII experiment performed a measurement of the energy-dependent Born cross section and the effective form factor for the $e^{+}e^{-}\to{\Omega}^{-}\bar{\Omega}^{+}$ reaction at the center-of-mass energies ranging from 3.4 to 4.7\thinspace GeV. The energy dependence of the dressed cross section is fitted with the hypothesis of a charmonium(-like) resonance [i.e., $\psi(3770)$, $\psi(4040)$, $\psi(4160)$, $Y(4230)$, $Y(4360)$, $\psi(4415)$, or $Y(4660)$] combined with a power-law function. The fit is applied to the recent BESIII data, combined with the earlier measurements from CLEO-c data, but no significance is found. The products of the branching fractions and the two-electronic partial widths for the assumed charmonium(-like) states decaying into the ${\Omega}^{-}\bar{\Omega}^{+}$ final state are also provided. In addition, by taking the world average value of the two-electronic partial width, the upper limits at the 90\% confidence level on the branching fractions of $\psi(3770)$, $\psi(4040)$, $\psi(4160)$, and $\psi(4415)$ decays into ${\Omega}^{-}\bar{\Omega}^{+}$ are determined for the first time. These are found to be at least an order of magnitude larger than expected from predictions using a scaling based on the observed electronic widths.
\end{abstract}


\end{frontmatter}

The study of the production of charmonium(-like) states above the open charm threshold in $e^{+}e^{-}$ annihilations plays a key role in testing the predictions of quantum chromodynamics (QCD)~\cite{bib0001,bib0002}. Although we have made many important achievements in both experiment and theory in understanding these charmonium-like states, the properties of some of the charmonium(-like) states are still not well established. In the past two decades, several vector states have been observed at the center-of-mass (c.m.) energies between 3.7 and 4.7\thinspace GeV at various $\EE$ colliders. Four charmonium(-like) states predicted by potential models~\cite{bib0001}, namely $\psi(3770)$, $\psi(4040)$, $\psi(4160)$, and $\psi(4415)$, have been observed as enhancements in the inclusive hadronic cross section~\cite{bib0003}. In addition, five new states such as $Y(4230)$, $Y(4260)$, $Y(4360)$, $X(4390)$, and $Y(4660)$ were reported using initial state radiation (ISR) processes $\EE\ar\gamma_{ISR}\pi^{+}\pi^{-}\jpsi (\psp)$ in the {\it BABAR}~\cite{bib0004,bib0005,bib0006,bib0007} and Belle experiments~\cite{bib0008,bib0009,bib0010,bib0011,bib0012}, or in energy scan experiments at the CLEO-c~\cite{bib0013} and BESIII~\cite{bib0014,bib0015,bib0016,bib0017,bib0018,bib0019,bib0066}.

The overpopulation of structures in this mass region and the mismatch of the properties between the potential model predictions and experimental measurements reflect our limited understanding of the strong interaction, particularly in its non-perturbative aspects.
Various models have been proposed to explain their nature~\cite{bib0001,bib0002,bib0020,bib0021,bib0022,bib0023}.
Among them, the production of light quark baryon-antibaryon ($\BB$) final states leads to a relatively simple topology, and the knowledge of the vector charmonium(-like) coupling to two-body baryonic final states would provide an additional way for understanding the properties of these states. In particular, charmless decays of these non-conventional states are proposed by the hybrid model~\cite{bib0020}. Meanwhile, if these states are considered pure charmonium~\cite{bib0022}, their baryonic decays, which have not yet been significantly observed, are strongly desired~\cite{bib0023}.
The\break understanding of the low mass puzzle of the $X(3872)$~\cite{bib0024,bib0025,bib0026} reminds us that studying the properties of charmonium(-like) states solely within the quenched quark model may be insufficient. In particular, above the open charm threshold, significant coupled channel effects are expected to arise in the unquenched picture~\cite{bib0022}, in which the physical states are no longer pure $c\bar{c}$ configurations.
Although numerous experimental studies~\cite{bib0027,bib0028,bib0029,bib0030,bib0031,bib0032,bib0033,bib0034,bib0035,bib0036,bib0037,bib0038,bib0039,bib0040,bib0041,bib0042,bib0043,bib0044,bib0045,bib0046,bib0047,bib0048,bib0049,bib0050,bib0051,bib0052,bib0053,bib0054,bib0055,bib0067,bib0068}
on the production of a pair of $\BB$ at the c.m. energies between 2 and 5\thinspace GeV have been performed by the BESIII and Belle experiments,
the experimental information on vector charmonium(-like) resonances decaying into the baryonic final states remains scarce, with only two evidences of $\psi(3770)\to \Lambda \bar{\Lambda}$ and $\Xi^{-}\bar{\Xi}^{+}$~\cite{bib0036,bib0047}.

In refs.~\cite{bib0056,bib0057}, the hadronic and leptonic decays of the $\psi(nS)$ states scale similarly according to perturbative QCD prediction with the principal quantum number $n$, i.e.,
\begin{equation}\label{BCS-11}
\frac{{\cal{B}}[\psi(n^{\prime}S)\to B\bar{B}]}{{\cal{B}}[\psi(nS) \to B\bar{B}]} = \frac{{\cal{B}}[\psi(n^{\prime}S)\to e^{+}e^{-}]}{{\cal{B}}[\psi(nS)\to e^{+}e^{-}]},
\end{equation}
with the assumption of non-resonance electromagnetic production domination of hadron pairs.
This assumed that the branching fractions of $\psi(3770)$, $\psi(4040)$, $\psi(4160)$, and $\psi(4415)$ to the $\BB$ final states scale with two-electronic partial widths when comparing to the $\psi(3686)$ state, {\it e.g.}, one estimates a negligible branching fraction for baryon octuplet and decuplet pairs as summarized in Table~\ref{tab:br}.
\begin{table}[!htbp]
    \caption{Branching fraction from the perturbative QCD prediction~\cite{bib0056,bib0057} for charmonium states decaying into the $B\bar{B}$ pair (unit in $\sim10^{-7}$), calculated based on the world average value of $\psi(3686) \to B\bar{B}$~\cite{bib0003}. 
    }
    \centering
    \scalebox{0.9}{
    \begin{tabular}{lcccc}\hline\hline
Decay   &$\psi(3770)$ &$\psi(4040)$ &$\psi(4160)$ &$\psi(4415)$ \\ \hline 
$p\bar{p}$                              &$3.55\pm0.30$      &$3.78\pm0.65$      &$2.55\pm1.23$      &$1.18\pm0.45$ \\
$n\bar{n}$                              &$3.70\pm0.34$      &$3.93\pm0.69$      &$2.66\pm1.28$      &$1.23\pm0.47$ \\
$\Lambda\bar{\Lambda}$                  &$4.61\pm0.39$      &$4.89\pm0.84$      &$3.31\pm1.59$      &$1.54\pm0.58$ \\
$\Sigma^+\bar\Sigma^-$                  &$2.94\pm0.26$      &$3.12\pm0.54$      &$2.11\pm1.02$      &$0.98\pm0.37$ \\
$\Sigma^0\bar\Sigma^0$                  &$2.84\pm0.25$      &$3.02\pm0.52$      &$2.04\pm0.98$      &$0.95\pm0.36$ \\
$\Sigma^-\bar\Sigma^+$                  &$3.41\pm0.29$      &$3.62\pm0.62$      &$2.45\pm1.18$      &$1.14\pm0.43$ \\
$\Sigma^{\ast\pm}\bar\Sigma^{\ast\mp}$  &$1.03\pm0.12$      &$1.09\pm0.21$      &$0.74\pm0.36$      &$0.34\pm0.13$ \\
$\Sigma^{\ast0}\bar\Sigma^{\ast0}$      &$0.83\pm0.11$      &$0.89\pm0.17$      &$0.60\pm0.29$      &$0.28\pm0.11$ \\
$\Xi^-\bar\Xi^+$                        &$3.47\pm0.30$      &$3.69\pm0.64$      &$2.49\pm1.20$      &$1.16\pm0.44$ \\
$\Xi^0\bar\Xi^0$                        &$2.78\pm0.53$      &$2.95\pm0.72$      &$2.00\pm1.02$      &$0.93\pm0.38$ \\
$\Xi^{\ast0}\bar\Xi^0$                  &$0.06\pm0.01$      &$0.07\pm0.01$      &$0.05\pm0.02$      &$0.02\pm0.01$ \\
$\Xi^{\ast-}\bar\Xi^{+}$                &$0.08\pm0.02$      &$0.09\pm0.02$      &$0.06\pm0.03$      &$0.03\pm0.01$\\
$\Xi^{\ast0}\bar\Xi^{\ast0}$            &$0.82\pm0.08$      &$0.87\pm0.16$      &$0.59\pm0.29$      &$0.27\pm0.10$ \\
$\Xi^{\ast-}\bar\Xi^{\ast+}$            &$1.39\pm0.14$      &$1.48\pm0.27$      &$1.00\pm0.48$      &$0.46\pm0.18$ \\
$\Omega^-\bar\Omega^+$                  &$0.68\pm0.06$      &$0.73\pm0.13$      &$0.49\pm0.24$      &$0.23\pm0.09$ \\
        \hline
        \hline
    \end{tabular}}
    \label{tab:br}
\end{table}
This means that the contributions of charmonium(-like) resonance decays are
negligibly small in all cases, and the effect of charmonium(-like) states can be safely attributed to electromagnetic production.
However, according to refs.~\cite{bib0036,bib0047}, the resulting branching fractions \allowbreak${\cal{B}}[\psi(3770)\to\Lambda\bar\Lambda] = 2.4^{+15.0}_{-2.0}(14.4^{+2.7}_{-14.0}) \times 10^{-5} $
and \allowbreak${\cal{B}}[\psi(3770)$\allowbreak$\to\Xi^-\bar\Xi^+] = (1.4 \pm 0.4) \times 10^{-4}$, 
are reported to be larger by at least an order of magnitude than the prediction based on a scaling of the two-electronic partial width using Eq. {(\ref{BCS-11})}. This implies that the contribution of charmonium(-like) states needs to be considered when interpreting the experimental data for the $\BB$ production cross section in $e^{+}e^{-}$ annihilation, and that these states may not be adequately described within a purely quenched picture.
Thus, more precise measurements of exclusive cross sections for baryonic final states above the open charm threshold are expected to validate the predictions of perturbative QCD.

In this article, we perform a fit to the dressed cross section of the $e^{+}e^{-}\to {\Omega}^{-}\bar{\Omega}^{+}$ reaction at the c.m. energies ($\sqrt{s}$) between 3.4 and 4.7\thinspace GeV. Possible charmonium(-like) states [i.e., $\psi(3770)$, $\psi(4040)$, $\psi(4160)$, $Y(4230)$, $Y(4360)$, $\psi(4415)$, and $Y(4660)$] decaying into the ${\Omega}^{-}\bar{\Omega}^{+}$ final state are searched for based on the experimental measurements, but no significant signal is found. Taking the world average value of the two-electronic partial width, the branching fractions for $\psi(3770)$, $\psi(4040)$, $\psi(4160)$ and $\psi(4415)$ decaying into the ${\Omega}^{-}\bar{\Omega}^{+}$ final state are also provided.

\section{Data collection}\label{Xsec1-1}
\begin{figure}[hbpt]
	\begin{center}
	\includegraphics[width=0.48\textwidth]{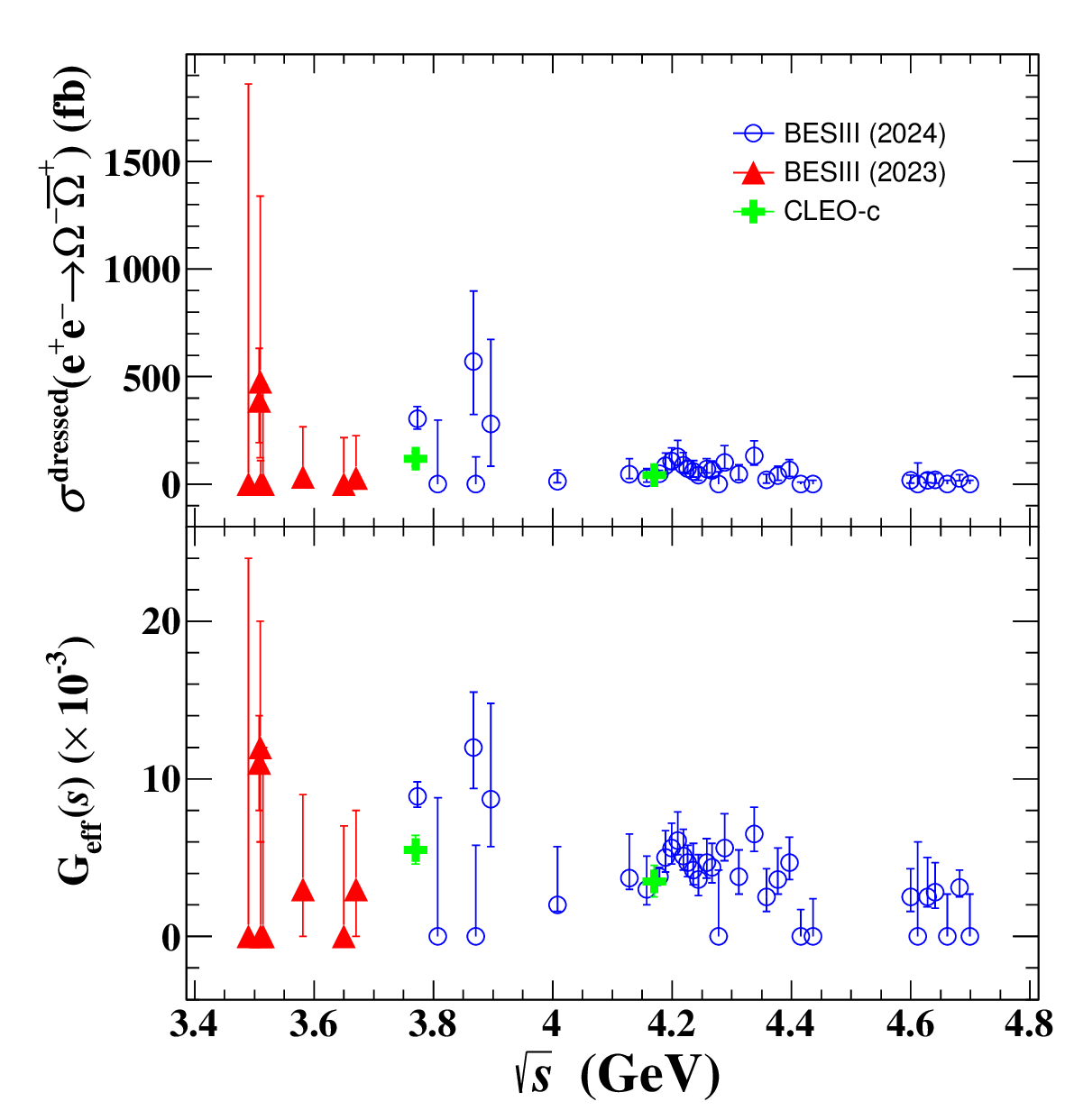}
	\end{center}
	\caption{The dressed cross section ($\sigma^{dressed}$) and ${\Omega}^{-}$ effective form factor ($G_{eff}(s)$) for the $ \EE\to \OOB$ reaction as a function of the c.m. energy from BESIII~\cite{bib0042,bib0061} and CLEO-c experiments~\cite{bib0057}. The error bars represent the statistical and systematic uncertainties summed in quadrature.
	}
	\label{Fig:XiXi::CS:BCS_VS_EFF}
\end{figure}

The ${\Omega}^{-}$ hyperon is a baryon containing three strange quarks, and its decays proceed only via the weak interaction and therefore have a relatively long lifetime. The spin of the ${\Omega}^{-}$ hyperon was predicted by the Eightfold Way and the quark model to be $J= 3/2$~\cite{bib0058}, which was subsequently confirmed by the model-dependent strategy in the {\it BABAR} experiment~\cite{bib0059}, and the model-independent strategy in the BESIII experiment~\cite{bib0060}.
Recently, the BESIII experiment reported an energy scan measurement of the Born cross section and effective form factor for the $\EE\to \OOB$ reaction using the data samples at c.m. energies between 3.4 and 4.7\thinspace GeV ~\cite{bib0042,bib0061}.
Previously, the CLEO-c experiment also presented a measurement of the hyperon pair production cross sections as well as the elastic and transition electromagnetic form factors for the $\EE\to \OOB$ reaction using the data in two c.m. energies at 3.77 and 4.17\thinspace GeV~\cite{bib0056,bib0057}. Both experiments employed a partial reconstruction method, i.e., reconstructing only one side (${\Omega}^{-}$ or $\bar{\Omega}^{+}$) based on the primary decay mode of ${\Omega}^{-}\to \Lambda K^{-}$ with $\Lambda \to p\pi^{-}$. The results of the above two measurements for the dressed cross section and the effective form factor are shown in Fig.~\ref{Fig:XiXi::CS:BCS_VS_EFF}.

\section{Fit to the cross section}\label{Xsec2-2}
The potential resonances in the line shape of the cross section for the $\EE\to \OOB$ reaction are searched for by fitting the dressed cross section including the vacuum polarization effect, using the least ${\chi}^{2}$ method
\begin{equation}
        \chi^{2} = \sum_{i=1}^{N}(\frac{Y_i-\eta_i}{\sigma_i})^2,
\label{Xeqn2-2}
\end{equation}
where $Y_i$ are the independent observed values, $\sigma_i$ are the corresponding uncertainties, and $\eta_i$ are the expected true values. If all $Y_i$ are correlated to each other, then it has
\begin{equation}
        \chi^{2} = \Delta X^{T}V^{-1}\Delta X.
\label{Xeqn3-3}
\end{equation}
This is done by considering the covariance matrix $V$ and the vector of residuals $\Delta X$ between the measured and fitted cross sections. The covariance matrix incorporates both correlated and uncorrelated uncertainties among different energy points. The systematic uncertainties associated with luminosity, event selection and reconstruction, and branching fraction are assumed to be fully correlated among the c.m. energies taken in refs.~\cite{bib0042,bib0057,bib0061}, while the other systematic uncertainties are assumed to be uncorrelated.
To account for the asymmetry of the statistical uncertainties, only the side closer to the fitted value is used to construct the covariance matrix.
With the uncertainties of the measurements of the two experiments’ treated as completely independent, the covariance matrix can be constructed as~\cite{bib0062}
\begin{equation}
    V = U + \sum_{k}b_k b_k^T + \sum_{k}c_k c_k^T,
\label{Xeqn4-4}
\end{equation}
where $U$ is the diagonal matrix of uncorrelated uncertainties and vector $b_k$ and $c_k$ encode the correlated systematic sources of the BESIII and CLEO-c experiments, respectively.

\begin{figure*}[]
	\begin{center}
        \includegraphics[width=0.42\textwidth]{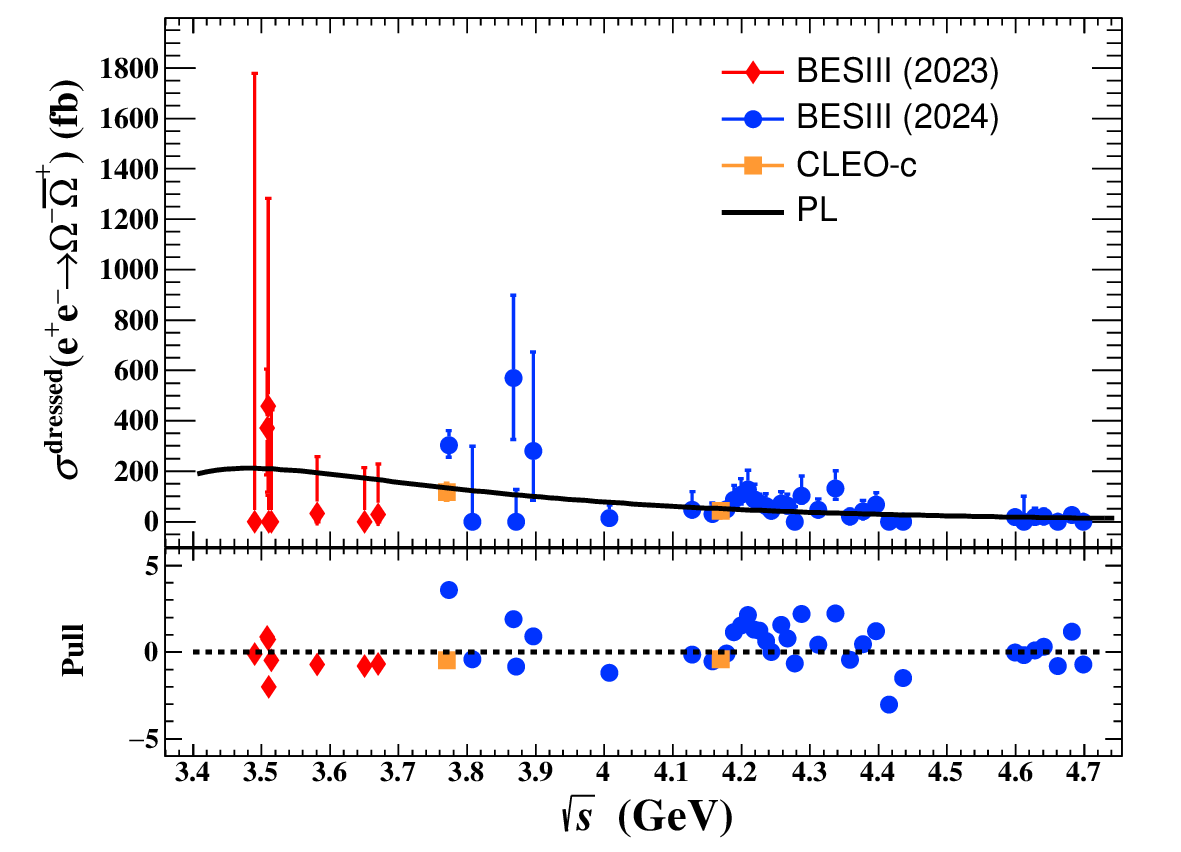}
        \includegraphics[width=0.42\textwidth]{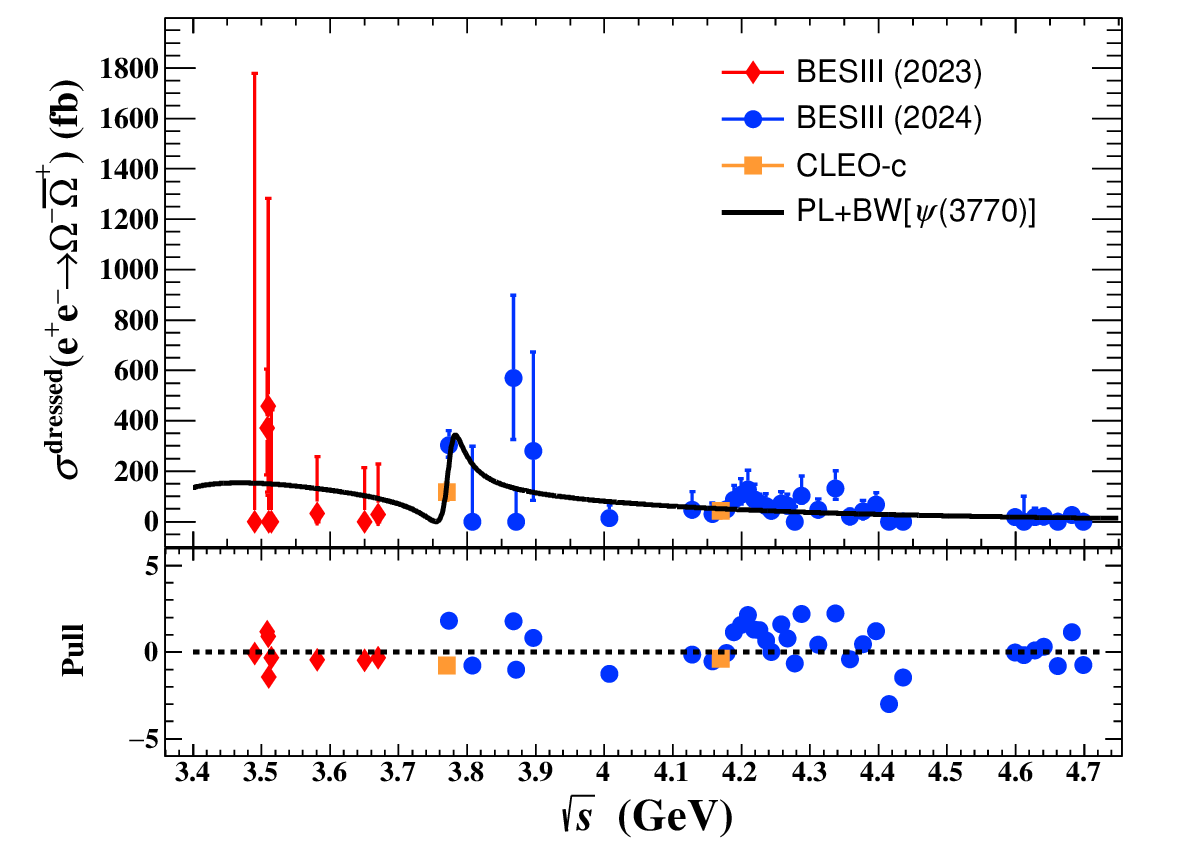}\\
        \includegraphics[width=0.42\textwidth]{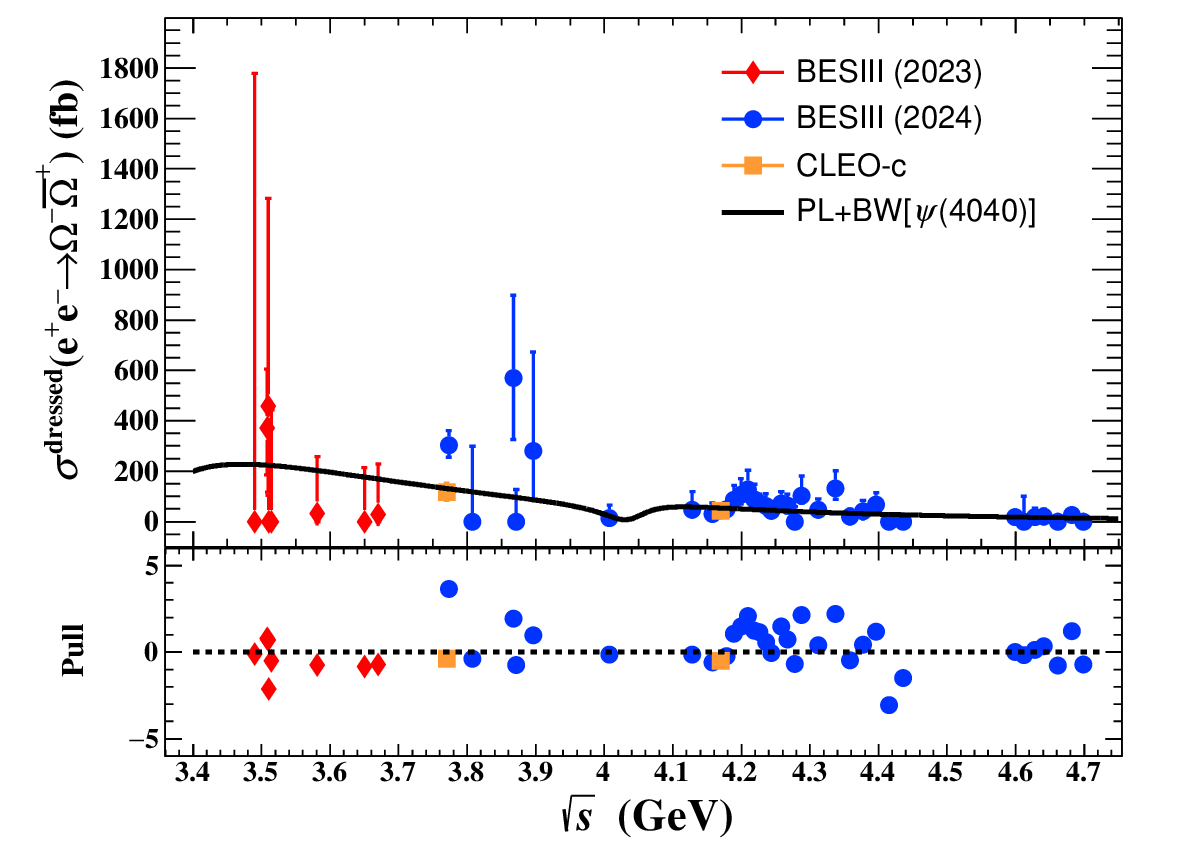}
        \includegraphics[width=0.42\textwidth]{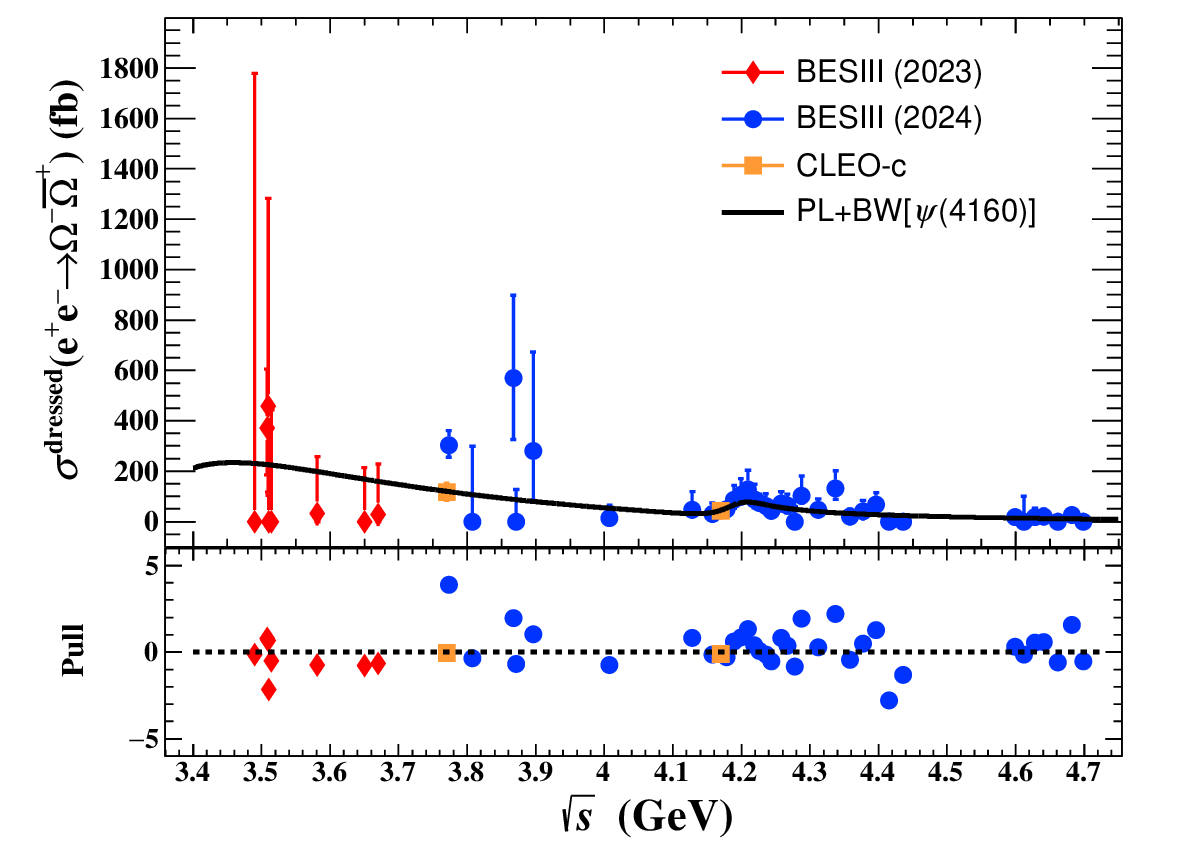}\\
        \includegraphics[width=0.42\textwidth]{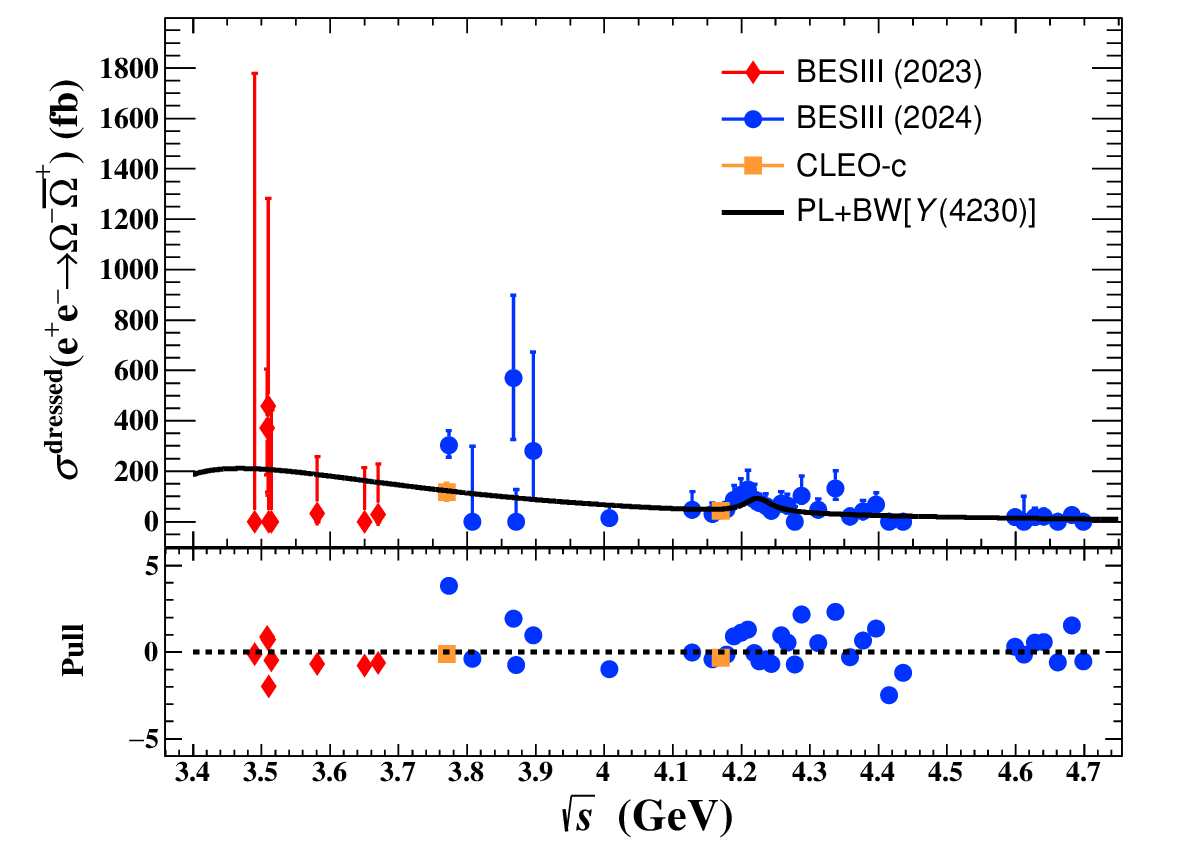}
        \includegraphics[width=0.42\textwidth]{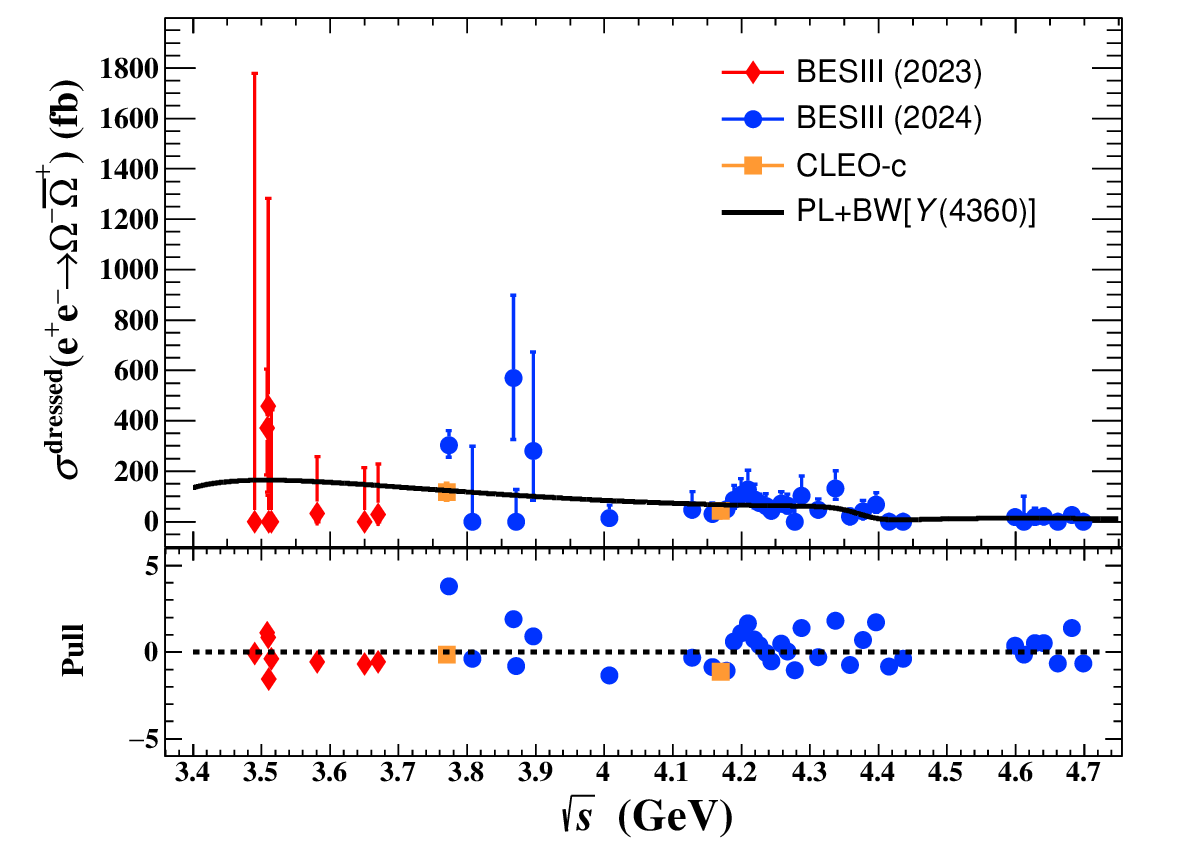}\\
        \includegraphics[width=0.42\textwidth]{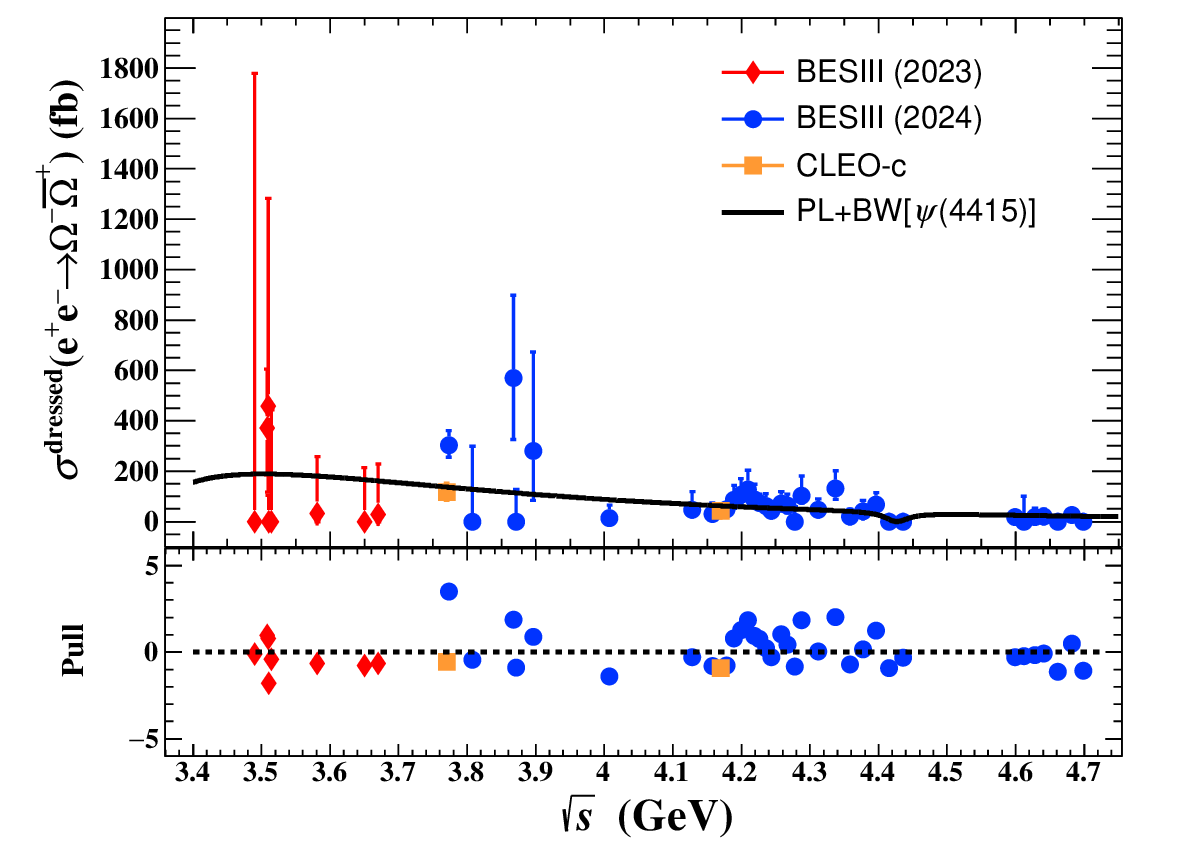}
        \includegraphics[width=0.42\textwidth]{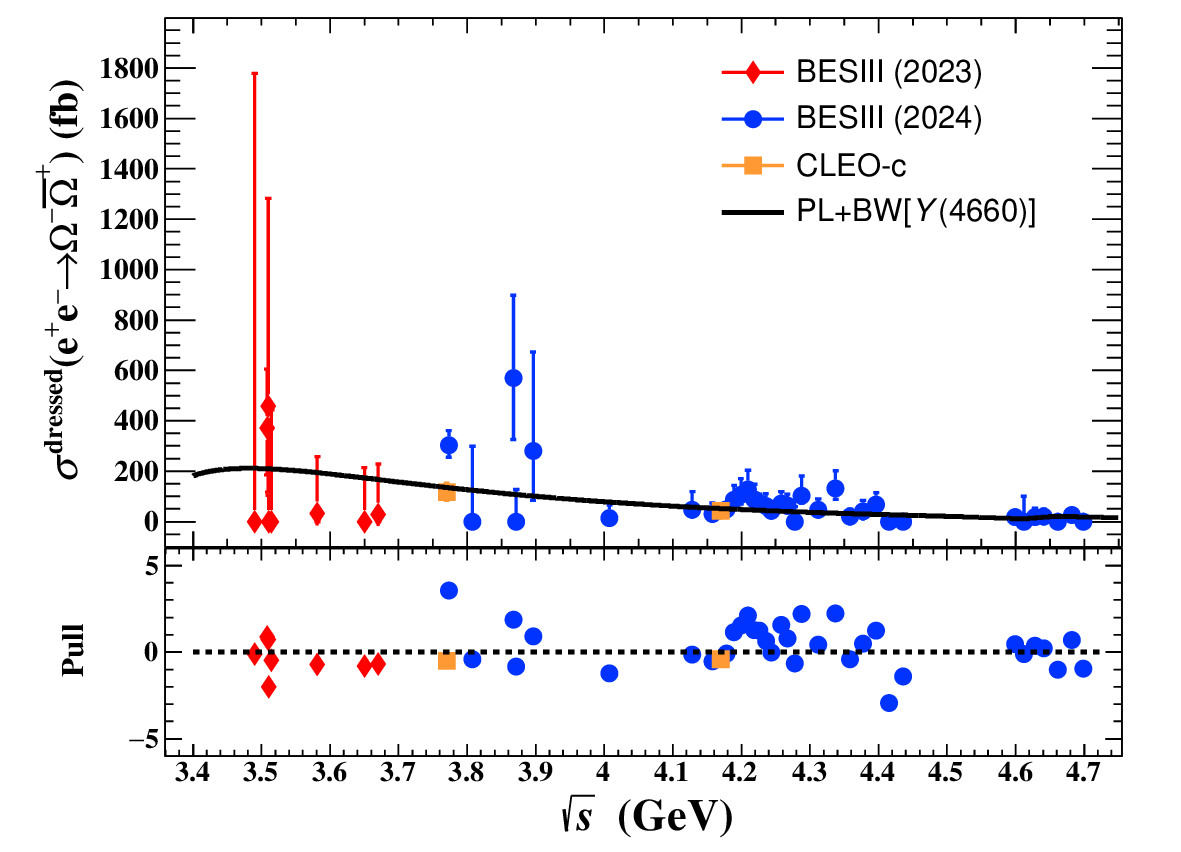}\\
	\end{center}
	\caption{Fit to the dressed cross section at the c.m. energies from 3.4 to 4.7 GeV with the assumptions of a PL function only (upper left) and a PL function plus a charmonium(-like) resonance [i.e., $\psi(3770)$, $\psi(4040)$, $\psi(4160)$, $Y(4230)$, $Y(4360)$, $\psi(4415)$, or $Y(4660)$]. Dots with error bars are the dressed cross sections, and the solid lines show the fit results. The error bars represent the statistical and systematic uncertainties summed in quadrature.}
	\label{Fig:XiXi::CS::Line-shape-3773}
\end {figure*}
\begin{figure*}[!htbp]
	\begin{center}
            \includegraphics[width=0.45\textwidth]{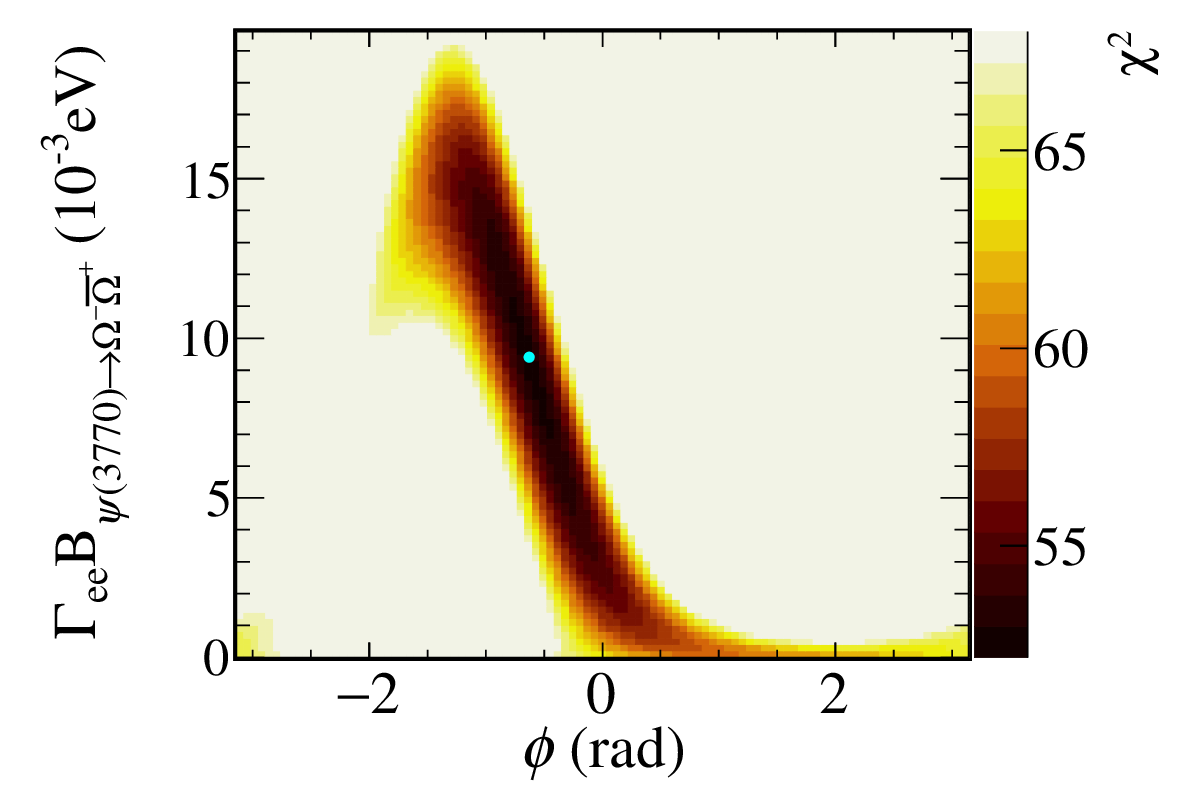}
            \includegraphics[width=0.45\textwidth]{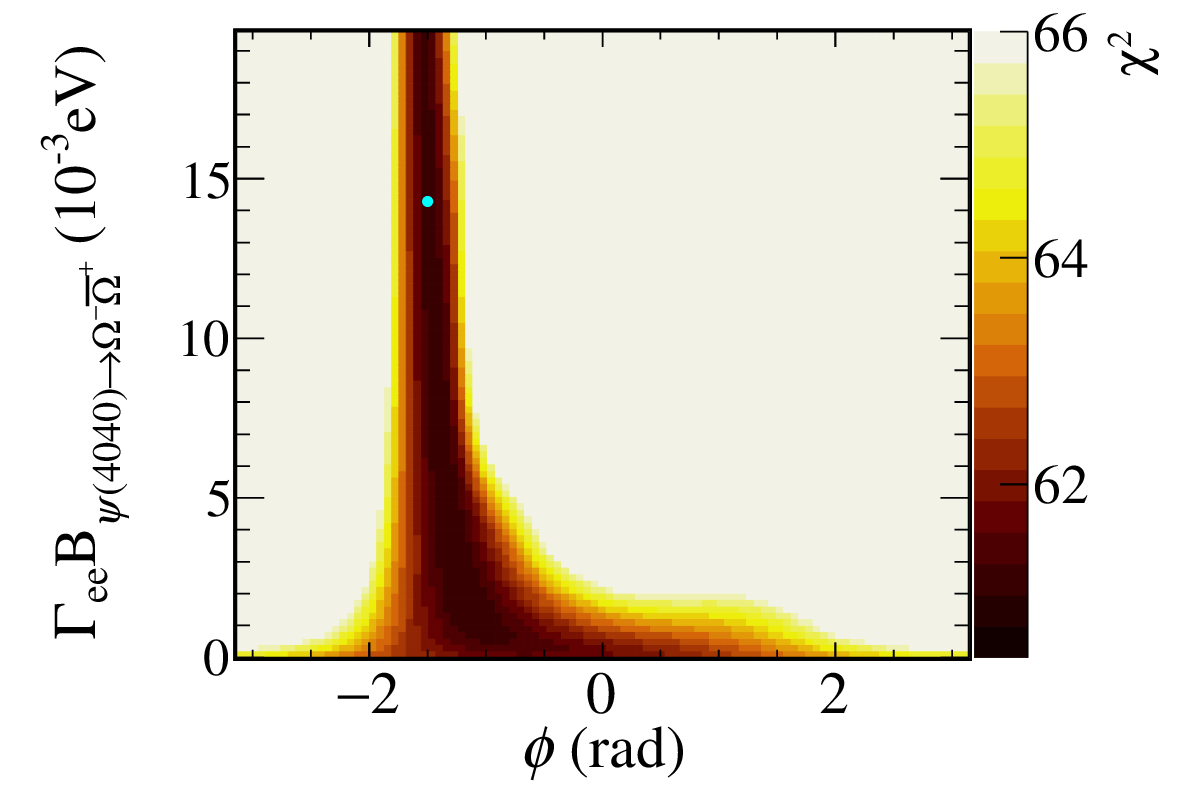}
            \includegraphics[width=0.45\textwidth]{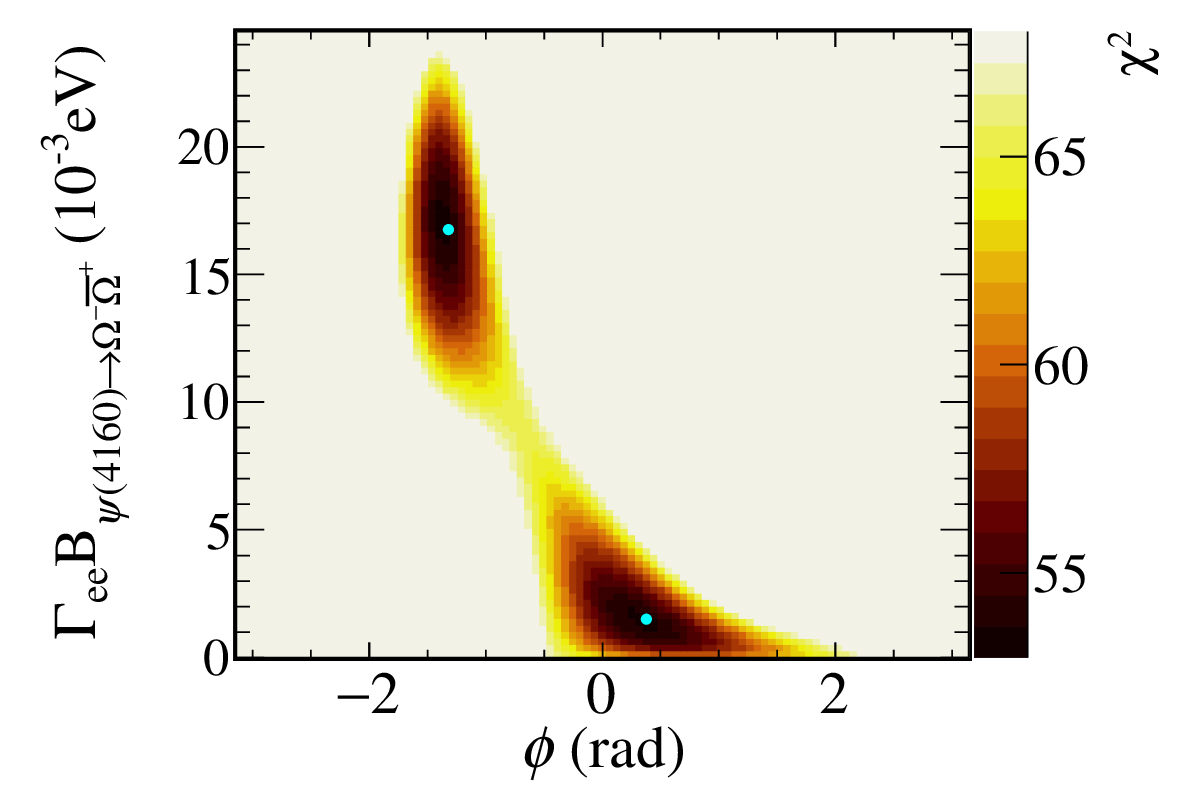}
            \includegraphics[width=0.45\textwidth]{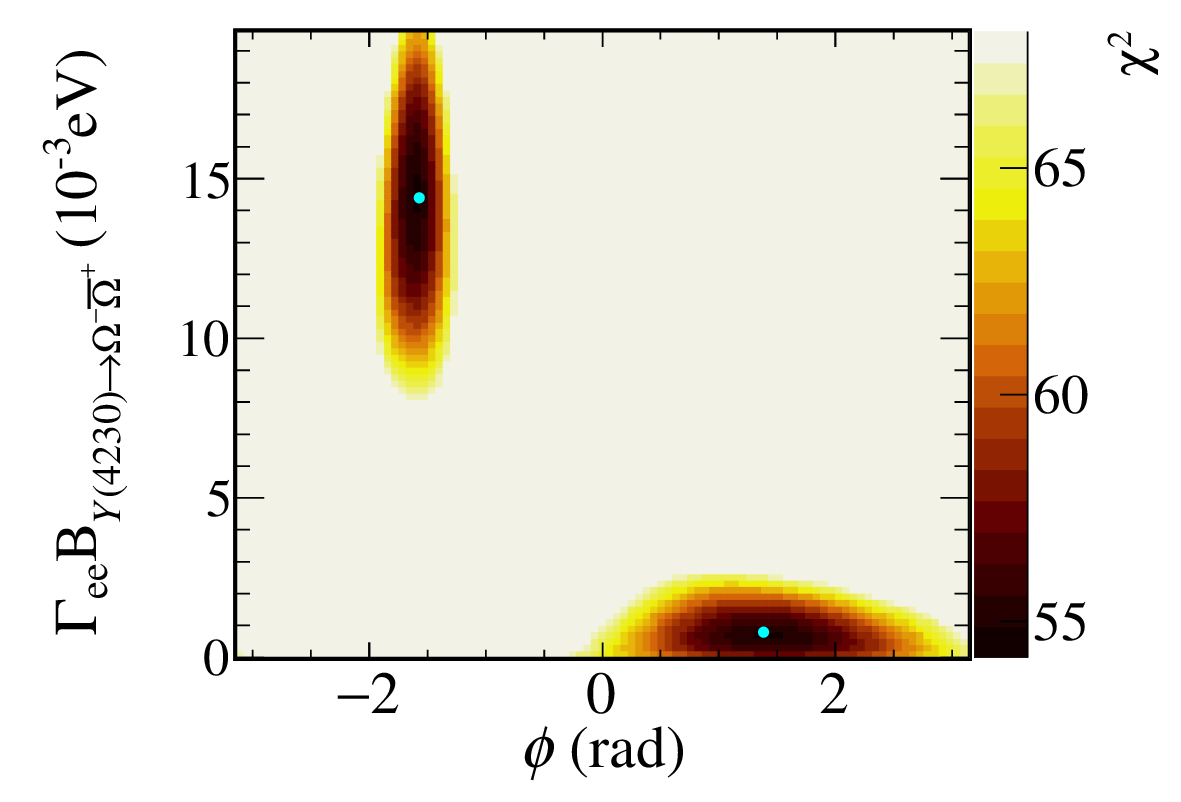}
            \includegraphics[width=0.45\textwidth]{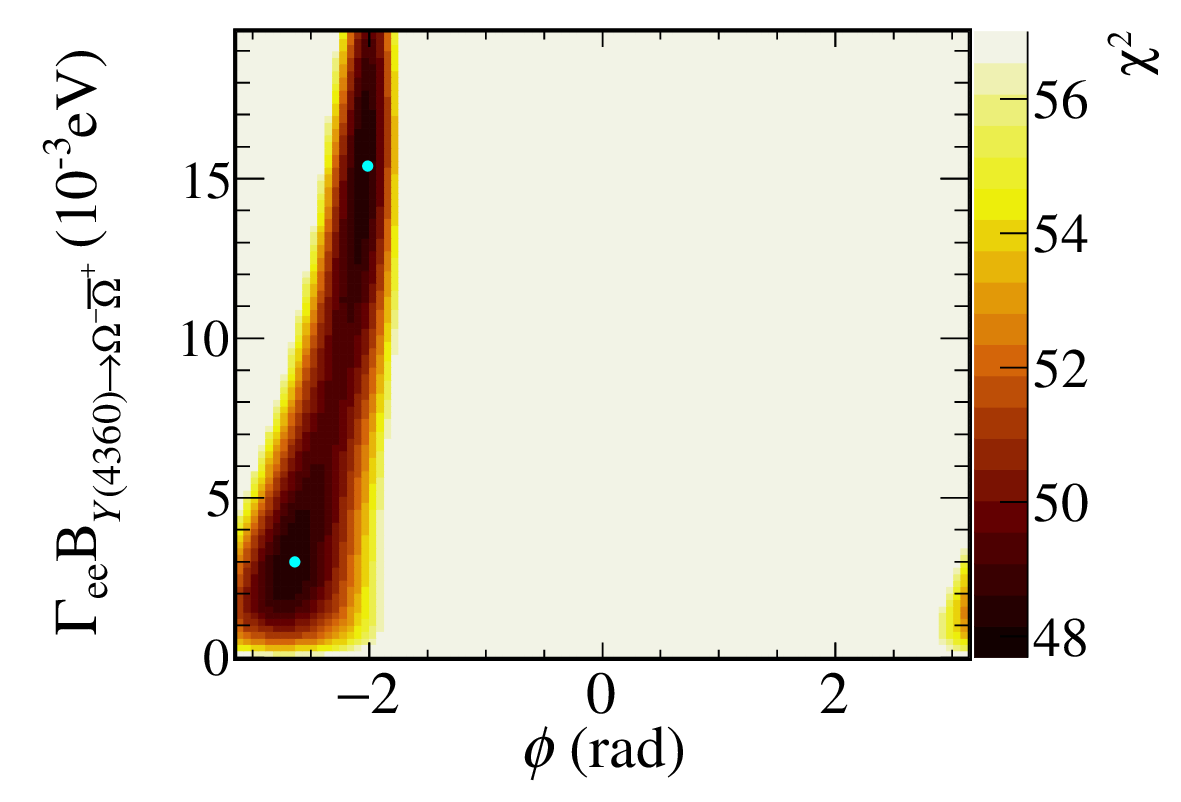}
            \includegraphics[width=0.45\textwidth]{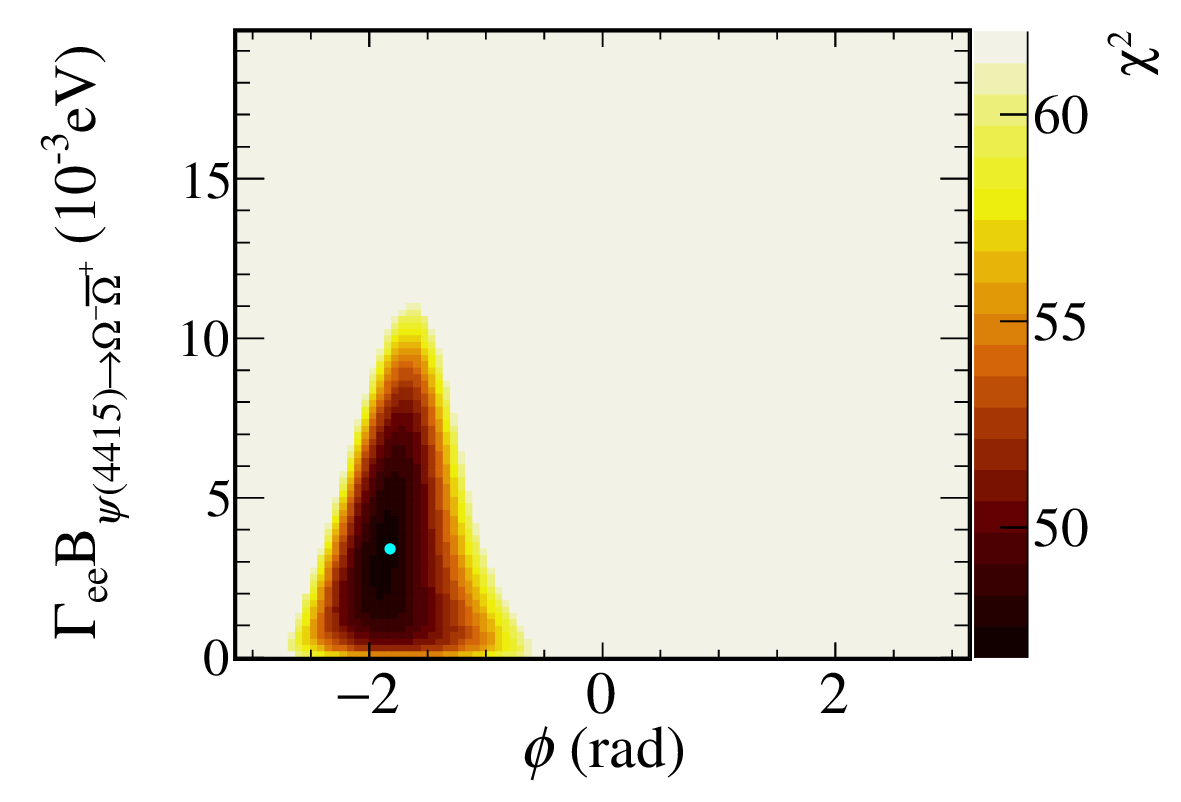}
            \includegraphics[width=0.45\textwidth]{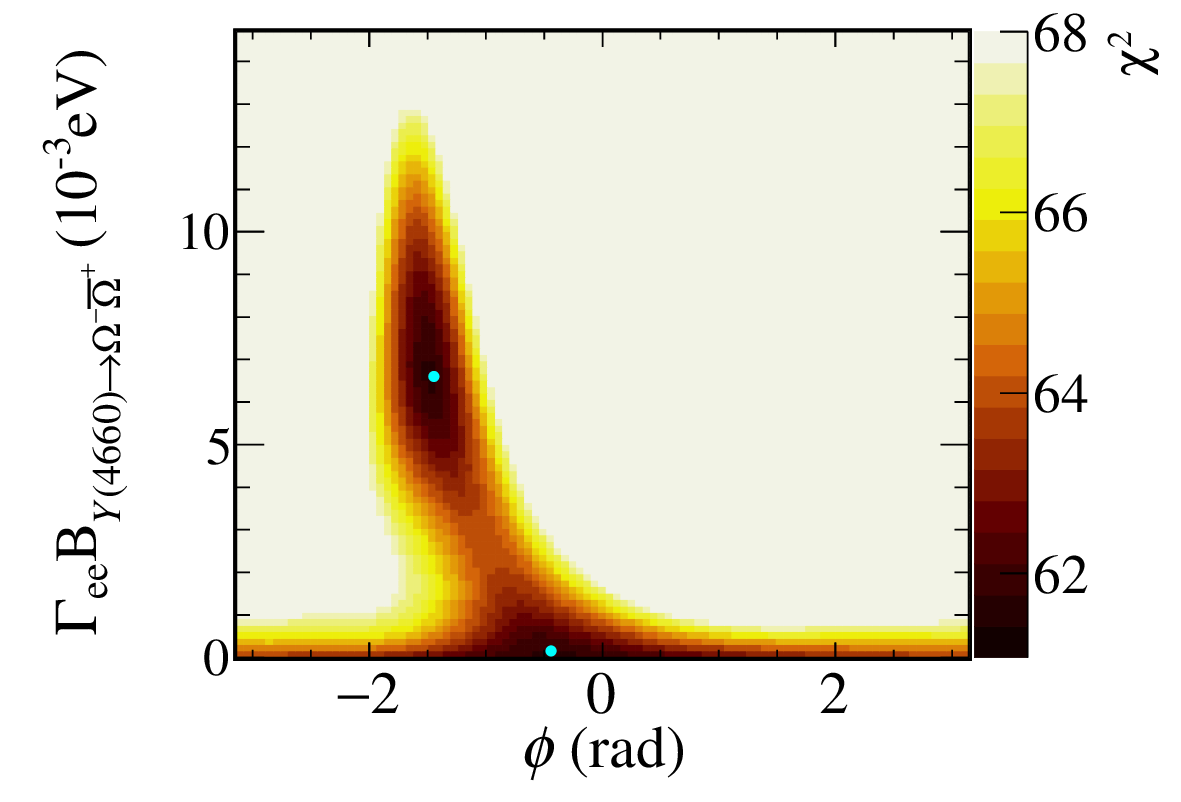}
	\end{center}
	\caption{The contour of $\Gamma_{ee}\mathcal{B}$ and $\phi$ on the distribution of $\chi^2$ values for $\psi(3770)$, $\psi(4040)$, $\psi(4160)$, $Y(4230)$, $Y(4360)$, $\psi(4415)$, or $Y(4660)$ decaying into the $\OOB$ final states. The cyan points represent the solutions from the best fit.}
	\label{Fig:scan}
\end {figure*}

\begin{table*}[!hbpt]
    \caption{The numerical results for the interference angle $\phi$, the products of the branching fraction and two-electronic partial width ($\Gamma_{ee}\mathcal{B}$)  and their upper limits ($\Gamma_{ee}\mathcal{B}_{UL}$)  for the possible charmonium(-like) states decaying into the $\Omega^-\bar{\Omega}^+$ final state to the dressed cross section of the $e^+e^-\to\Omega^-\bar{\Omega}^+$ reaction. Here for other charmonium(-like) states $Y(4230)$, $Y(4360)$ and $Y(4660)$, their branching fraction has not been determined due to lack of two-electronic partial width measurement. The first uncertainty is from statistical one and second one is from alternative fit with a continuum contribution plus seven charmonium(-like) resonances.
    }
    \centering
    \renewcommand{\arraystretch}{1.5}
    \begin{tabular}{l@{\hspace{2em}} l@{\hspace{4em}} l@{\hspace{4em}} l@{\hspace{4em}} l} 
    \hline 
    \hline
        Resonance        &$~~~~~\phi~(\rm{rad})$       &$\Gamma_{ee}\mathcal{B}~(<\Gamma_{ee}\mathcal{B}_{UL})~(10^{-3}\rm{eV})$  &$\mathcal{B}~(<\mathcal{B}_{UL})~(10^{-6})$     &$\chi^2/n.d.f$\\ \hline
        $\psi(3770)$                &$-0.6 \pm 0.3$      &~~$9  \pm 4  \pm 1~(<16)$      &$35 \pm 16~(<61)$         &$51/(44-4)$                        \\ \hline    
        $\psi(4040)$                &$-1.5 \pm 0.2$      &~~$14 \pm 17 \pm 9~(<35)$      &$14 \pm 20~(<41)$         &$59/(44-4)$                        \\ \hline    
        $\psi(4160)^{1}$            &$~~~0.4  \pm 0.2$   &~~$1   \pm 1  \pm 7$           &$3  \pm 2$                &\multirow{2}{*}{$52/(44-4)$}       \\
        $\psi(4160)^{2}$            &$-1.4 \pm 0.1$      &~~$17 \pm 1  \pm 9~(<20)$      &$36 \pm 17~(<42)$                                              \\ \hline          
        $Y(4230)^{1}$               &$~~~1.4  \pm 0.3$   &~~$1  \pm 1  \pm 2$            &~~~~~~~~$-$               &\multirow{2}{*}{$53/(44-4)$}       \\
        $Y(4230)^{2}$               &$-1.6 \pm 0.1$      &~~$15 \pm 1  \pm 12~(<18)$     &~~~~~~~~$-$                                                   \\ \hline          
        $Y(4360)^{1}$               &$-2.6 \pm 0.2$      &~~$3  \pm 2  \pm 2$            &~~~~~~~~$-$               &\multirow{2}{*}{$46/(44-4)$}       \\
        $Y(4360)^{2}$               &$-2.0 \pm 0.1$      &~~$15 \pm 3  \pm 14~(<21)$     &~~~~~~~~$-$                                                   \\ \hline          
        $\psi(4415)$                &$-2.0 \pm 0.2$      &~~$6  \pm 4  \pm 1~(<14)$      &$17  \pm 13~(<40)$        &$45/(44-4)$                        \\ \hline     
        $Y(4660)^{1}$               &$-0.5 \pm 0.6$      &~~$1  \pm 1  \pm 5$            &~~~~~~~~$-$               &\multirow{2}{*}{$59/(44-4)$}       \\
        $Y(4660)^{2}$               &$-1.5 \pm 0.1$      &~~$5  \pm 1  \pm 1~(<9)$       &~~~~~~~~$-$                                                   \\ \hline \hline
    \end{tabular}
    \label{tab:multisolution}
\end{table*}

Assuming the cross section of $\EE\ar\OOB$
includes a charmonium(-like) resonance [i.e., $\psi(3770)$, $\psi(4040)$, $\psi(4160)$,
$Y(4230)$, $Y(4360)$, $\psi(4415)$, or $Y(4660)$]
plus a continuum contribution, a fit to the dressed
cross section with the coherent sum of a power-law (PL)
function plus a Breit-Wigner (BW) function
\begin{equation}\label{BCS-1}
	\sigma^{\rm dressed}(\sqrt{s})= {\left|c_{0}\frac{\sqrt{P(\sqrt{s})}}{\sqrt{s}^{n}} + e^{i\phi}{\rm BW}(\sqrt{s})\sqrt{\frac{P(\sqrt{s})}{P(M)}}\right|}^{2},
\end{equation}
is applied. Here $\phi$ is the relative phase between the BW function
 \begin{equation}
{\rm BW}(\sqrt{s}) =\frac{\sqrt{12\pi\Gamma_{ee}{\cal{B}}\Gamma}}{s-M^{2}+iM\Gamma},
\label{Xeqn6-6}
\end{equation}
and the PL function, $c_0$ and $n$ are free fit parameters,
\begin{equation}
    \sqrt{P(\sqrt{s})} = \frac{\sqrt{(s-4M^2)s}}{2\sqrt{s}}
\label{Xeqn7-7}
\end{equation}
 is the two body phase space factor, the mass $M$ and the total width $\Gamma$ are fixed to the assumed resonance with the PDG values~\cite{bib0003}, and $\Gamma_{ee}{\cal{B}}$ is the product of the two-electronic partial width and the branching fraction for the assumed resonance decaying into the $\OOB$ final state. Fig.~\ref{Fig:XiXi::CS::Line-shape-3773} shows the fit to the dressed cross section assuming the resonance [i.e., $\psi(3770)$, $\psi(4040)$, $\psi(4160)$, $Y(4230)$, $Y(4360)$, $\psi(4415)$, or $Y(4660)$] and without resonance assumption.
 Note that the $Y(4500)$~\cite{bib0063} is not included in this analysis, as there are no data points in its corresponding energy region.
 The parameters without resonance assumption are fitted to be ($c_0 = 0.1 \pm 0.1, n = 6.7 \pm 0.4$) with the fit goodness of $\chi^{2}/n.d.f = 60/(36-2)$, and other parameters with resonance assumption are summarized in Table~\ref{tab:multisolution}. Taking into account systematic uncertainties, the significance for each resonance is calculated by comparing the change of $\chi^{2}/n.d.f$ with and without the resonance assumption.  For different assumptions of charmonium(-like) states, $\psi(3770)$, $\psi(4040)$, $\psi(4160)$, $Y(4230)$, $Y(4360)$, $\psi(4415)$, or $Y(4660)$, are all fitted by Eq. {(\ref{BCS-1})} one at a time. The significance is also evaluated to be less than 3$\sigma$ for all assumed charmonium(-like) states.
The corresponding $\Gamma_{ee}{\cal B}$ values for these charmonium(-like) states decays into the ${\Omega}^{-}\bar{\Omega}^{+}$ final state are extracted from the fits.
Due to the quadratic form of the cross section like Eq. {(\ref{BCS-1})}, there are multiple solutions~\cite{bib0065}, which can be determined by scanning the parameters $\phi$ and $\Gamma_{ee}\mathcal{B}$, similar to the method used in ref.~\cite{bib0036}. Fig.~\ref{Fig:scan} shows the contour of $\Gamma_{ee}\mathcal{B}$ and $\phi$ on the distribution of $\chi^2$ values for each set of parameters. The results of the fit and their multiple solutions are summarized in Table~\ref{tab:multisolution}.
It should be noted that for certain resonance states, such as $\psi(3770)$, $\psi(4040)$, and $\psi(4415)$, the two solutions are so close that they cannot be differentiated at the current level of precision. As a result, the contour shows only a single minimum. Consequently, only this minimum value is reported.
The upper limits of $\Gamma_{ee}\mathcal{B}$ at the 90\% C.L. are determined from the change in the $\chi^2$ value with respect to the nominal fit~\cite{bib0064}. Specifically, the upper limit is set by requiring
\begin{equation}
    \Delta\chi^2 = \chi^2 - \chi^2_{\min} = \Delta\chi^2_{90\%}(\Delta n_{\rm df}),
\label{Xeqn8-8}
\end{equation}
where $\Delta n_{\rm df}$ is the number of degrees of freedom, and $\Delta\chi^2_{90\%}$ is the critical value of the $\chi^2$ distribution corresponding to 90\% probability.
After that, by taking the world average value of the two-electronic partial width, the branching fractions for $\psi(3770)$, $\psi(4040)$, $\psi(4160)$, $\psi(4415)$ decaying into ${\Omega}^{-}\bar{\Omega}^{+}$ final state at 90\% C.L. are also extracted to be
\begin{center}
${\cal{B}}[\psi(3770)\to\Omega^-\bar\Omega^+] < 6\times10^{-5}$,
${\cal{B}}[\psi(4040)\to\Omega^-\bar\Omega^+] < 4\times10^{-5}$,
${\cal{B}}[\psi(4160)\to\Omega^-\bar\Omega^+] < 4\times10^{-5}$,
${\cal{B}}[\psi(4415)\to\Omega^-\bar\Omega^+] < 4\times10^{-5}$,
\end{center}
which are found to be at least an order of magnitude larger than expected from the predictions using a scaling based on the observed electronic widths~\cite{bib0056,bib0057}.

\begin{table*}[!hbpt]
    \caption{Summary of the ranges of $\Gamma_{ee}\mathcal{B}$ values from the alternative fit.
    }
    \centering
    \renewcommand{\arraystretch}{1.5}
    \begin{tabular}{c c c c c c c c}
    \hline
    \hline
        Resonance                                       &~~~$\psi(3770)$~~     &~~$\psi(4040)$~~       &~~$\psi(4160)$~~       &~~$Y(4230)$~~      &~~$Y(4360)$~~      &~~$\psi(4415)$~~       &~~$Y(4660)$   \\ \hline
        $\Gamma_{ee}\mathcal{B} (10^{-3}~{\rm eV})$     &~~~[9,~10]~~          &~~[4,~9]~~           &~~[5,~34]~~            &~~[2,~18]~~        &~~[0,~102]~~        &~~[2,~88]~~            &~~[5,~7]       \\
        \hline
        \hline
    \end{tabular}
    \label{tab:multisolution2}
\end{table*}

\begin{figure}[!hbpt]
	\begin{center}
        \includegraphics[width=0.49\textwidth]{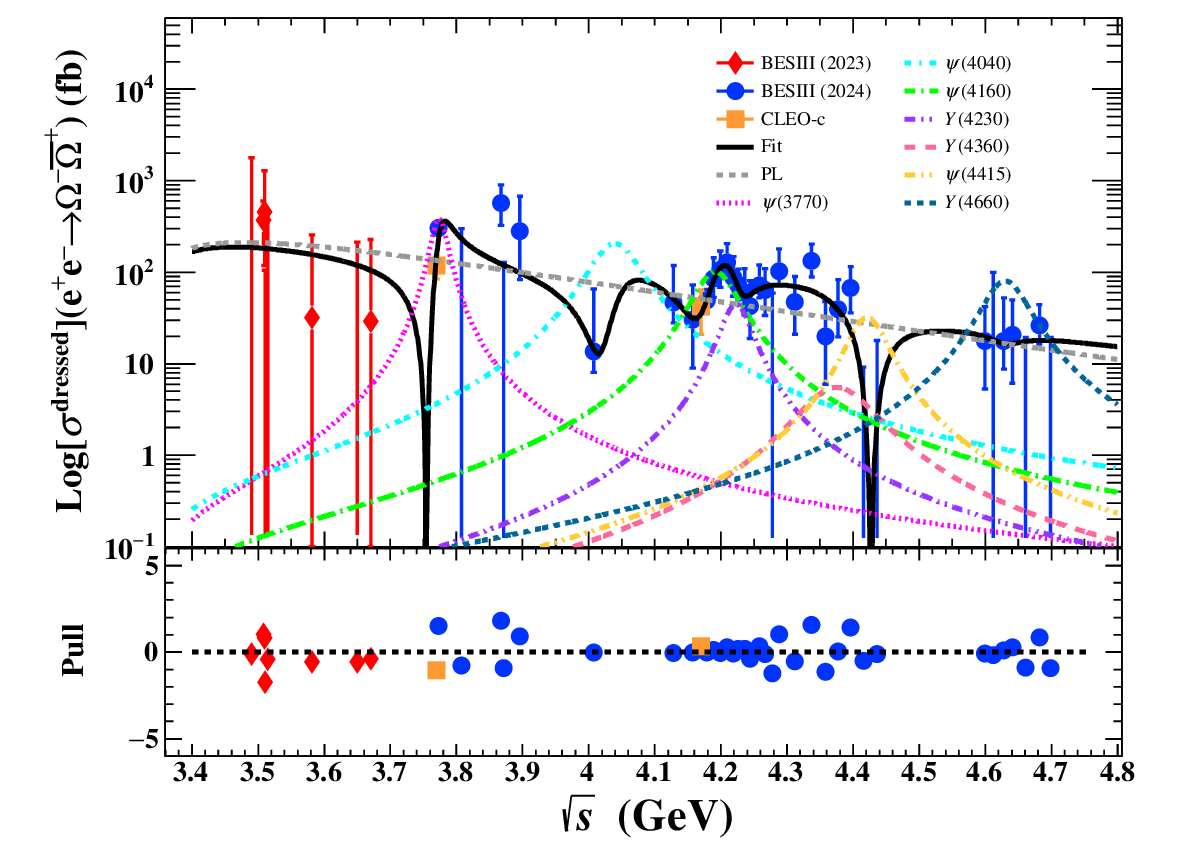}
	\end{center}
	\caption{Fit to the dressed cross section at the c.m. energies from 3.4 to 4.7 GeV with the assumption of a continuum contribution (PL) plus seven charmonium(-like) resonances [i.e., $\psi(3770)$, $\psi(4040)$, $\psi(4160)$, $Y(4230)$, $Y(4360)$, $\psi(4415)$, or $Y(4660)$]. Dots with error bars are the dressed cross sections, the solid lines show the fit results, and different dashed lines represent the continuum contribution and different resonance states.}
	\label{Fig:XiXi::CS::Line-shape-3773:sum}
\end {figure}

Additionally, an alternative fit to the dressed cross section with the coherent sum of a continuum contribution plus seven charmonium(-like) resonances is shown in Fig.~\ref{Fig:XiXi::CS::Line-shape-3773:sum}, and the corresponding fit results are summarized in Table~\ref{tab:multisolution2}. The fitted baseline values of $\Gamma_{ee}\mathcal{B}$ are consistent with those determined when introducing one resonance at a time. The differences are taken as systematic uncertainties (Table~\ref{tab:multisolution}).

\section{Summary}\label{Xsec3-3}

In summary, we perform a search for charmonium(-like) states decaying into the ${\Omega}^{-}\bar{\Omega}^{+}$ final states based on the recent measurement of the energy-dependent cross section and the effective form factor for the $e^{+}e^{-}\to{\Omega}^{-}\bar{\Omega}^{+}$ reaction at the c.m. energies between 3.4 and 4.7\thinspace GeV by the BESIII experiment. A fit to the energy dependence of the dressed cross section is performed with the assumption of a charmonium(-like) resonance [i.e., $\psi(3770)$, $\psi(4040)$, $\psi(4160)$, $Y(4230)$, $Y(4360)$, $\psi(4415)$, or $Y(4660)$] plus a PL function to the data from the recent BESIII measurements~\cite{bib0042,bib0061} combined additionally with two energy points from the previous CLEO-c measurements~\cite{bib0056,bib0057}. The significance is also evaluated for these assumed charmonium(-like) states, and no significance is found.
And the products of branching fractions and two-electronic partial widths for these assumed charmonium(-like) states decaying into the ${\Omega}^{-}\bar{\Omega}^{+}$ final state are also provided.
In addition, by taking the world average value of the two-electronic partial width, the branching fractions for $\psi(3770)$, $\psi(4040)$, $\psi(4160)$, $\psi(4415)$ decays into ${\Omega}^{-}\bar{\Omega}^{+}$ final state at 90\% C.L. are also determined, which are found to be at least an order of magnitude larger than expected from predictions using a scaling based on observed electronic widths.
This tendency is consistent with some previous measurements~\cite{bib0036,bib0047,bib0050,bib0051,bib0053}, suggesting that the $B\bar{B}$ production might not originate purely from electromagnetic processes, but could involve contributions from charmonium(-like) states, which may be more consistent with the unquenched picture.
With the current statistics, this result indicates the importance of the contribution of vector charmonium(-like) states when interpreting the experimental data in terms of, e.g., electromagnetic structure observables, and provides more useful information for understanding the charmonium(-like) states coupling to the $B\bar{B}$ final states.

\section*{Acknowledgements}

This work was supported by 
the Fundamental Research Funds for the Central Universities Nos.
lzujbky-2025-it06, lzujbky-2025-ytA05, lzujbky-2024-jdzx06;
the Natural Science Foundation of Gansu Province No. 22JR5RA389, No.25JRRA799;
the ‘111 Center’ under Grant No. B20063;
the National Natural Science Foundation of China under Contract No. 12247101.


\begin{thebibliography}{68}
\bibitem{bib0001} \href{https://doi.org/10.1140/epjc/s10052-010-1534-9}{N. Brambilla, et al., Eur. Phys. J. C \textbf{71} (2011) 1534.}

\bibitem{bib0002} \href{https://doi.org/10.1088/1674-1137/40/4/042001}{R. A. Briceno, et al., Chin. Phys. C \textbf{40} (2016) 042001.}

\bibitem{bib0003} \href{https://doi.org/10.1103/PhysRevD.110.030001}{S. Navas, et al., Phys. Rev. D \textbf{110} (2024) 030001.}

\bibitem{bib0004} \href{https://doi.org/10.1103/PhysRevLett.95.142001}{B. Aubert, et al., BaBar Collaboration, Phys. Rev. Lett. \textbf{95} (2005) 142001.}

\bibitem{bib0005} \href{https://doi.org/10.1103/PhysRevLett.98.212001}{B. Aubert, et al., BaBar Collaboration, Phys. Rev. Lett. \textbf{98} (2007) 212001.}

\bibitem{bib0006} \href{https://doi.org/10.1103/PhysRevD.86.051102}{J. P. Lees, et al., BaBar Collaboration, Phys. Rev. D \textbf{86} (2012) 051102.}

\bibitem{bib0007} \href{https://doi.org/10.1103/PhysRevD.89.111103}{J. P. Lees, et al., BaBar Collaboration, Phys. Rev. D \textbf{89} (2014) 111103.}

\bibitem{bib0008} \href{https://doi.org/10.1103/PhysRevLett.99.182004}{C. Z. Yuan, et al., Belle Collaboration, Phys. Rev. Lett. \textbf{99} (2007) 182004.}

\bibitem{bib0009} \href{https://doi.org/10.1103/PhysRevLett.99.142002}{X. L. Wang, et al., Belle Collaboration, Phys. Rev. Lett. \textbf{99} (2007) 142002.}

\bibitem{bib0010} \href{https://doi.org/10.1103/PhysRevLett.101.172001}{G. Pakhlova, et al., Belle Collaboration, Phys. Rev. Lett. \textbf{101} (2008) 172001.}

\bibitem{bib0011} \href{https://doi.org/10.1103/PhysRevLett.110.252002}{Z. Q. Liu, et al., Belle Collaboration, Phys. Rev. Lett. \textbf{110} (2013) 252002.}

\bibitem{bib0012} \href{https://doi.org/10.1103/PhysRevD.91.112007}{X. L. Wang, et al., Belle Collaboration, Phys. Rev. D \textbf{91} (2015) 112007.}

\bibitem{bib0013} \href{https://doi.org/10.1103/PhysRevLett.96.162003}{T. E. Coan, et al., CLEO Collaboration, Phys. Rev. Lett. \textbf{96} (2006) 162003.}

\bibitem{bib0014} \href{https://doi.org/10.1103/PhysRevLett.114.092003}{M. Ablikim, et al., BESIII Collaboration, Phys. Rev. Lett. \textbf{114} (2015) 092003.}

\bibitem{bib0015} \href{https://doi.org/10.1103/PhysRevLett.118.092001}{M. Ablikim, et al., BESIII Collaboration, Phys. Rev. Lett. \textbf{118} (2017) 092001.}

\bibitem{bib0016} \href{https://doi.org/10.1103/PhysRevD.97.052001}{M. Ablikim, et al., BESIII Collaboration, Phys. Rev. D \textbf{97} (2018) 052001.}

\bibitem{bib0017} \href{https://doi.org/10.1103/PhysRevD.101.012008}{M. Ablikim, et al., BESIII Collaboration, Phys. Rev. D \textbf{101} (2020) 012008.}

\bibitem{bib0018} \href{https://doi.org/10.1103/PhysRevD.102.012009}{M. Ablikim, et al., BESIII Collaboration, Phys. Rev. D \textbf{102} (2020) 012009.}

\bibitem{bib0019} \href{https://doi.org/10.1103/PhysRevD.104.052012}{M. Ablikim, et al., BESIII Collaboration, Phys. Rev. D \textbf{104} (2021) 052012.}

\bibitem{bib0066} \href{https://doi.org/10.1103/PhysRevLett.133.081901}{M. Ablikim, et al., BESIII Collaboration, Phys. Rev. Lett. \textbf{133} (2024) 081901.}

\bibitem{bib0020} \href{https://doi.org/10.1016/j.physletb.2005.09.016}{F. E. Close, P. R. Page, Phys. Lett. B \textbf{628} (2005) 215.}

\bibitem{bib0021} \href{https://doi.org/10.1016/j.physrep.2016.05.004}{H. X. Chen, W. Chen, X. L. Liu, S. L. Zhu, Phys. Rept. \textbf{639} (2016).}

\bibitem{bib0022} \href{https://doi.org/10.1103/PhysRevD.99.114003}{J. Z. Wang, D. Y. Chen, X. L. Liu, T. Matsuki, Phys. Rev. D \textbf{99} (2019) 114003.}

\bibitem{bib0023} \href{https://doi.org/10.1016/j.physletb.2022.137292}{R. Q. Qian, Q. Huang, X. Liu, Phys. Lett. B \textbf{833} (2022) 137292.}

\bibitem{bib0024} \href{https://doi.org/10.1103/PhysRevD.72.034010}{Y. S. Kalashnikova, Phys. Rev. D \textbf{72} (2005) 034010.}

\bibitem{bib0025} \href{https://doi.org/10.1016/j.physletb.2009.09.033}{O. Zhang, C. Meng, H. Q. Zheng, Phys. Lett. B \textbf{680} (2009) 453.}

\bibitem{bib0026} \href{https://doi.org/10.1103/PhysRevD.79.094004}{B. Q. Li, K. T. Chao, Phys. Rev. D \textbf{79} (2009) 094004.}

\bibitem{bib0027} \href{https://doi.org/10.1103/PhysRevD.87.112011}{M. Ablikim, et al., BESIII Collaboration, Phys. Rev. D \textbf{87} (2013) 112011.}

\bibitem{bib0028} \href{https://doi.org/10.1103/PhysRevD.93.072003}{M. Ablikim, et al., BESIII Collaboration, Phys. Rev. D \textbf{93} (2016) 072003.}

\bibitem{bib0029} \href{https://doi.org/10.1016/j.physletb.2017.04.037}{M. Ablikim, et al., BESIII Collaboration, Phys. Lett. B \textbf{770} (2017) 217.}

\bibitem{bib0030} \href{https://doi.org/10.1103/PhysRevLett.120.132001}{M. Ablikim, et al., BESIII Collaboration, Phys. Rev. Lett. \textbf{120} (2018) 132001.}

\bibitem{bib0031} \href{https://doi.org/10.1016/j.nuclphysb.2019.04.012}{X. F. Wang, B. Li, Y. N. Gao, X. Lou, Nucl. Phys. B \textbf{941} (2019) 861.}

\bibitem{bib0032} \href{https://doi.org/10.1103/PhysRevD.100.051101}{M. Ablikim, et al., BESIII Collaboration, Phys. Rev. D \textbf{100} (2019) 051101.}

\bibitem{bib0033} \href{https://doi.org/10.1103/PhysRevLett.124.032002}{M. Ablikim, et al., BESIII Collaboration, Phys. Rev. Lett. \textbf{124} (2020) 032002.}

\bibitem{bib0034} \href{https://doi.org/10.1103/PhysRevD.103.012005}{M. Ablikim, et al., BESIII Collaboration, Phys. Rev. D \textbf{103} (2021) 012005.}

\bibitem{bib0035} \href{https://doi.org/10.1016/j.physletb.2021.136557}{M. Ablikim, et al., BESIII Collaboration, Phys. Lett. B \textbf{820} (2021) 136557.}

\bibitem{bib0036} \href{https://doi.org/10.1103/PhysRevD.104.L091104}{M. Ablikim, et al., BESIII Collaboration, Phys. Rev. D \textbf{104} (2021) L091104.}

\bibitem{bib0037} \href{https://doi.org/10.1103/PhysRevD.104.092012}{M. Ablikim, et al., BESIII Collaboration, Phys. Rev. D \textbf{104} (2021) 092012.}

\bibitem{bib0038} \href{https://doi.org/10.1103/PhysRevD.105.L011101}{M. Ablikim, et al., BESIII Collaboration, Phys. Rev. D \textbf{105} (2022) L011101.}

\bibitem{bib0039} \href{https://doi.org/10.3390/sym14010065}{X. F. Wang, G. S. Huang, Symmetry \textbf{14} (2022) 65.}

\bibitem{bib0040} \href{https://doi.org/10.1007/JHEP06(2022)074}{M. Ablikim, et al., BESIII Collaboration, JHEP \textbf{06} (2022) 074.}

\bibitem{bib0041} \href{https://doi.org/10.1103/PhysRevD.106.L091101}{M. Ablikim, et al., BESIII Collaboration, Phys. Rev. D \textbf{106} (2022) L091101.}

\bibitem{bib0042} \href{https://doi.org/10.1103/PhysRevD.107.052003}{M. Ablikim, et al., BESIII Collaboration, Phys. Rev. D \textbf{107} (2023) 052003.}

\bibitem{bib0043} \href{https://doi.org/10.1103/PhysRevD.108.L011101}{M. Ablikim, et al., BESIII Collaboration, Phys. Rev. D \textbf{108} (2023) L011101.}

\bibitem{bib0044} \href{https://doi.org/10.3390/sym15010214}{H. Liu, J. Zhang, X. Wang, Symmetry \textbf{15} (2023) 214.}

\bibitem{bib0045} \href{https://doi.org/10.1007/JHEP10(2023)081}{M. Ablikim, et al., BESIII Collaboration, JHEP \textbf{10} (2023) 081.}

\bibitem{bib0046} \href{https://doi.org/10.1103/PhysRevLett.131.191901}{M. Ablikim, et al., BESIII Collaboration, Phys. Rev. Lett. \textbf{131} (2023) 191901.}

\bibitem{bib0047} \href{https://doi.org/10.1007/JHEP11(2023)228}{M. Ablikim, et al., BESIII Collaboration, JHEP \textbf{11} (2023) 228.}

\bibitem{bib0048} \href{https://doi.org/10.1007/JHEP07(2024)258}{M. Ablikim, et al., BESIII Collaboration, JHEP \textbf{07} (2024) 258.}

\bibitem{bib0049} \href{https://doi.org/10.1007/JHEP12(2024)186}{M. Ablikim, et al., BESIII Collaboration, JHEP \textbf{12} (2024) 186.}

\bibitem{bib0050} \href{https://doi.org/10.1007/JHEP11(2024)062}{M. Ablikim, et al., BESIII Collaboration, JHEP \textbf{11} (2024) 062.}

\bibitem{bib0051} \href{https://doi.org/10.1007/JHEP05(2024)022}{M. Ablikim, et al., BESIII Collaboration, JHEP \textbf{05} (2024) 022.}

\bibitem{bib0052} \href{https://pos.sissa.it/465/087}{X. F. Wang, PoS \textbf{465} (2025) 087.}

\bibitem{bib0053} \href{https://doi.org/10.1103/PhysRevD.111.L051502}{M. Ablikim, et al., BESIII Collaboration, Phys. Rev. D \textbf{111} (2025) L051502.}

\bibitem{bib0054} \href{https://arxiv.org/abs/2502.15117}{X. F. Wang, X. Liu, Y. N. Gao, arXiv:2502.15117.}

\bibitem{bib0055} \href{https://arxiv.org/abs/2504.05584}{M. Ablikim, et al., BESIII Collaboration, arXiv:2504.05584.}

\bibitem{bib0067} \href{https://www.mdpi.com/2073-8994/18/1/200}{R. Y. Zhang, X. F. Wang, Symmetry \textbf{18} (2026) 200.}

\bibitem{bib0068} \href{https://arxiv.org/abs/2602.23835}{M. Ablikim, et al., BESIII Collaboration, arXiv:2602.23835.}

\bibitem{bib0056} \href{https://doi.org/10.1016/j.physletb.2014.10.025}{S. Dobbs, A. Tomaradze, T. Xiao, K. K. Seth, G. Bonvicini, Phys. Lett. B \textbf{739} (2014) 90.}

\bibitem{bib0057} \href{https://doi.org/10.1103/PhysRevD.96.092004}{S. Dobbs, K. K. Seth, A. Tomaradze, T. Xiao, G. Bonvicini, Phys. Rev. D \textbf{96} (2017) 092004.}

\bibitem{bib0058} \href{https://doi.org/10.1103/PhysRev.125.1067}{M. Gell-Mann, Phys. Rev. \textbf{125} (1962) 1067.}

\bibitem{bib0059} \href{https://doi.org/10.1103/PhysRevLett.97.112001}{B. Aubert, et al., BaBar Collaboration, Phys. Rev. Lett. \textbf{97} (2006) 112001.}

\bibitem{bib0060} \href{https://doi.org/10.1103/PhysRevLett.126.092002}{M. Ablikim, et al., BESIII Collaboration, Phys. Rev. Lett. \textbf{126} (2021) 092002.}

\bibitem{bib0061} \href{https://arxiv.org/abs/2508.01359}{M. Ablikim, et al., BESIII Collaboration, arXiv:2508.01359.}

\bibitem{bib0062} \href{https://doi.org/10.1140/epjc/s10052-022-10581-w}{D. Britzger, Eur. Phys. J. C \textbf{82} (2022) 731.}

\bibitem{bib0063} \href{https://doi.org/10.1088/1674-1137/ac945c}{M. Ablikim, et al., BESIII Collaboration, Chin. Phys. C \textbf{46} (2022) 111002.}

\bibitem{bib0065} \href{https://doi.org/10.1103/PhysRevD.99.072007}{Y. Bai, D. Y. Chen, Phys. Rev. D \textbf{99} (2019) 072007.}

\bibitem{bib0064} \href{https://doi.org/10.1088/1674-1137/32/5/007}{Y. S. Zhu, Chin. Phys. C \textbf{32} (2008) 363.}


\end{thebibliography}
\end{document}